\documentclass{article}
\usepackage{amsfonts}
\usepackage{bm}

\begin{document}

\author{A. B\'{e}rard$^{a}$, J. Lages$^{b}$, and H. Mohrbach$^{c}$
\\
\\
\textit{$^{a}$ L.P.M.C.T., Institut\ de\ Physique, } \\
\textit{1 blvd. Fran\c{c}ois Arago, 57070, Metz, France} \\
\\
\textit{$^b$ Ames Lab, Department of Physics,} \\
\textit{Iowa State University, Ames 50011-3020 IA, USA} \\
\\
\textit{$^c$ Institut Charles Sadron, CNRS UPR 022,} \\
\textit{6 rue Boussingault, 67083 Strasbourg Cedex, France}}

\title{Angular Symmetry Breaking Induced by Electromagnetic Field}
\maketitle

\begin{abstract}
It is well known that velocities does not commute in presence of an
electromagnetic field. This property implies that angular algebra
symmetries, such as the sO(3) and Lorentz algebra symmetries, are broken. To
restore these angular symmetries we show the necessity of adding the
Poincar\'{e} momentum $\mathbf{M}$\ to the simple angular momentum
$\mathbf{L}$.
These restorations performed succesively in a flat space and in a
curved space lead in each cases to the generation of a Dirac magnetic
monopole. In the particular case of the Lorentz algebra we consider an
application of our theory to the gravitoelectromagnetism. In this last case
we establish a qualitative relation giving the mass spectrum for dyons.
\end{abstract}

\section{Introduction}

The concept of symmetry breaking is fundamental in science and particularly
in physics. In this work we will focus our attention on the effects of the
noncommutativity on the angular algebra symmetries such as the sO(3) algebra
symmetry and the Lorentz algebra symmetry. Theories in noncommutative
geometry has been at the center of recent interest \cite{CONNES,SEIBERG}.
Instead of the noncommutativity between coordinates $\mathbf{x}=\left\{
x^{i}\right\} _{i=1,N}$  here we will rather use noncommutativity between
velocities $\dot{\mathbf{x}}=\left\{ \dot{x}^{i}\right\} _{i=1,N}$ leaving
the study of the more general case where neither coordinates nor velocities
commute for another paper \cite{NOUS5}. As we will see in the following in
the tangent bundle space $T(\mathcal{M})$ of a manifold $\mathcal{M}$
endowed with a Poisson structure the noncommutativity between velocities
implies a gauge curvature $F^{ij}(\mathbf{x})$, ie. an electromagnetic
f\/ield. This gauge curvature breaks explicitly the angular algebra symmetry
of the dynamical system. The particular case of the breaking of the Lorentz
algebra in presence of a covariant Hamiltonian was already investigated in a
recent paper \cite{NOUS2}. Here we embedded the formalism used in \cite
{NOUS2} in a more general one in def\/{}ining a Poisson structure where the
dynamics are governed by a covariant Hamiltonian. We also show that the
brackets def\/{}ining the Poisson structure on $T(\mathcal{M})$ are
equivalent (at least up to the second order) to Moyal brackets def\/{}ined
on the tangent bundle space $T(\mathcal{M})$ where the noncommutative
parameters are related to the electromagnetic f\/ield.

Contrary to the standard approach used commonly in the study of the gauge
theories, we do not settle our formalism in the cotangent bundle space $%
T^{*}(\mathcal{M})$, ie. the space of coordinates $\left\{ \mathbf{x},%
\mathbf{p}\right\} $, but we continue the prospection of the gauge theories
in the tangent bundle space $T(\mathcal{M})$, ie. the space of coordinates $%
\left\{ \mathbf{x},\dot{\mathbf{x}}\right\} $. The main reason for this
choice is that the generalized momentum $\mathbf{p}$ is not gauge invariant
and this is particularly important when we consider angular algebra
symmetry. Indeed, the Lie algebra is always trivially realized when the
angular momentum $\mathbf{L}$ is expressed in terms of the generalized
momentum $\mathbf{p}$. In the tangent bundle space $T(\mathcal{M})$ we show
that the angular algebra symmetry is broken by electromagnetic f\/ield
issued from the noncommutativity between velocities. This angular algebra
symmetry is restored by the introduction of the Poincar\'{e} momentum \cite
{POINCARE} which implies the existence of a Dirac magnetic monopole \cite
{DIRAC,NOUS4,NOUS1}. Here we insist further than in \cite{NOUS4} on the link
between the restoration of the angular Lie algebra symmetries and the
generation of magnetic monopoles. Note that a similar monopole also arises
from the electromagnetic U(1) gauge theory when one require the dual
symmetry under rotation of the electric and magnetic f\/ields in the free
Maxwell equations. In our case we do not need the dual symmetry (excepted
in Section \ref{Lorentz} where we use the Hodge duality for other purpose)
since the monopole f\/ield naturally arises from the restoration of the
angular algebra symmetry which was broken by the electromagnetic f\/ield. In
this paper we perform the restoration for the sO(3) algebra symmetry
succesively in a f\/{}lat and in a curved space and then for the local
Lorentz algebra symmetry.  F\/{}inally, at the end of the paper, using the
Hodge duality on our theory and the recent work of Mashhoon \cite{MASHHOON}
we develop an application to the gravitoelectromagnetism theory.

We would like to remark that the Moyal brackets we use in this paper are
closely connected to those introduced by Feynman in his remarkable
demonstration of the Maxwell equations where he tried to develop a
quantization procedure without the resort of a Lagrangian or a Hamiltonian.
Feynman' s ideas were exposed by Dyson in an elegant publication \cite{DYSON}.
The interpretation of the Feynman' s derivation of the Maxwell
equations has aroused a great interest among physicists. In particular
Tanimura \cite{TANIMURA} has generalized the Feynman' s derivation in a
Lorentz covariant form with a scalar time evolution parameter. An extension
of the Tanimura' s approach has been achieved \cite{NOUS3} in using the
Hodge duality to derive the two groups of Maxwell equations with a
magnetic monopole in f\/{}lat and curved spaces. In \cite{LEE} the
descriptions of relativistic and non relativistic particles in an
electromagnetic f\/ield was studied, whereas in \cite{CHOU} a dynamical
equation for spinning particles was proposed. A rigorous mathematical
interpretation of Feynman' s derivation connected to the inverse problem for
the Poisson dynamic has been formulated in \cite{CARINENA}. Also the papers 
\cite{HOJMAN} and \cite{HUGHES} considered the Feynman' s derivation in the
frame of the Helmholtz' s inverse problem for the calculus of variations.
More recently, some works \cite{MONTESINOS}, \cite{SINGH}, \cite{SILAGADZE}
have provided new looks on the Feynman' s derivation of the Maxwell
equations. One can mention a tentative of extension of Feynman' s derivation
of the Maxwell equations to the case of noncommutative geometry using the
standard Moyal brackets \cite{BOULAHOUAL}.

The paper is organized as follows. In Section \ref{formalism} we introduce
the formalism used throughout the paper. The dynamical equation will be
given by a covariant Hamiltonian def\/{}ined as in Goldstein' s textbook 
\cite{GOLDSTEIN} and by deformed Poisson brackets. We show these deformed
Poisson brackets can be def\/{}ined as the second order approximation of
generalized Moyal brackets. In Section \ref{so3sec} we recall how to restore
the sO(3) algebra symmetry in the case of a f\/{}lat space \cite{NOUS2} and
we extend the restoration to the case of a curved space. In Section \ref
{lorentz} we restore the Lorentz algebra symmetry in a curved space and we
propose an application of our formalism to the case of the
gravitoelectromagnetism theory. We then derive a qualitative relation giving
the mass spectrum for dyons. In Section \ref{conclusion} we summarize the
main achievements of this work.

\section{Mathematical foundations}

\label{formalism}

Let $\mathcal{M}$ be a $N$-dimensional vectorial manifold diffeomorphic to $%
\mathbb{R}^{N}$. Let a particle with a mass $m$ and an electrical charge $q$ be
described by the vector $\mathbf{x}=\{x^{i}\}_{i=1,\dots ,N}$ which
def\/{}ines its position on the manifold $\mathcal{M}$. Let $\tau $ be the
parameter of the group of diffeomorphisms $\mathcal{G}:\mathbb{R}\times 
\mathcal{M}\rightarrow \mathcal{M}$ such as $\mathcal{G}(\tau ,\mathbf{x})=%
\mathcal{G}^{\tau }\mathbf{x}=\mathbf{x}(\tau )$. Then taking $\tau $ as the
time parameter of our physical system we are able to def\/{}ine a velocity
vector $\dot{\mathbf{x}}\in \mathcal{M}$ as $\dot{\mathbf{x}}=\displaystyle%
\frac{d\mathbf{x}}{d\tau }=\mathcal{G}^{\tau }\mathbf{x}=\{\dot{x}^{i}(\tau
)\}_{i=1,\dots ,N}$. Let $T(\mathcal{M})$ be the tangent bundle space
associated with the manifold $\mathcal{M}$, a point on $T(\mathcal{M})$ is
described then by a $2N$ dimensional vector $\mathbf{X}=\{\mathbf{x},\mathbf{%
\dot{x}}\}$.

\subsection{Poisson structure}

Let $A^{0}(T(\mathcal{M}))=C^{\infty }(T(\mathcal{M}),\mathbb{R})$ be the
algebra of differential functions def\/{}ined on the manifold $T(\mathcal{M}%
) $. We def\/{}ine a Poisson structure on $T(\mathcal{M})$ which is an
internal skew-symmetric bilinear multiplicative law on $A^{0}(T(\mathcal{M}%
)) $ denoted $(f,g)\rightarrow \lbrack f,g]$ and satisfying the Leibnitz
rule 
\begin{equation}
\lbrack f,gh]=[f,g]h+[f,h]g  \label{Leibnitz}
\end{equation}
and the Jacobi identity 
\begin{equation}
J(f,g,h)=[f,[g,h]]+[g,[h,f]]+[h,[f,g]]=0.  \label{Jacobi}
\end{equation}
The manifold $T(\mathcal{M})$ with such a Poisson structure is called a
Poisson manifold. We def\/{}ine a dynamical system on the Poisson manifold $%
T(\mathcal{M})$ by the following differential equation 
\begin{equation}
\frac{df}{d\tau }=[f,H]  \label{dynamic}
\end{equation}
where $H\in A^{0}(T(\mathcal{M}))$ is the Hamiltonian of the dynamical
system.

With such def\/{}initions we derive the following important relations for
functions belonging to $A^{0}(T(\mathcal{M}))$ 
\begin{eqnarray}
\left[ f(\mathbf{X}),h(\mathbf{X})\right] &=&\left\{ f(\mathbf{X}),h(\mathbf{%
X})\right\}  \nonumber \\
&+&\left[ x^{i},x^{j}\right] \frac{\partial f(\mathbf{X})}{\partial x^{i}}%
\frac{\partial h(\mathbf{X})}{\partial x^{j}}  \nonumber \\
&+&\left[ \dot{x}^{i},\dot{x}^{j}\right] \frac{\partial f(\mathbf{X})}{%
\partial \dot{x}^{i}}\frac{\partial h(\mathbf{X})}{\partial \dot{x}^{j}},
\label{bracket1}
\end{eqnarray}
where we have introduced Poisson-like brackets def\/{}ined by 
\begin{equation}
\left\{ f(\mathbf{X}),g(\mathbf{X})\right\} =\left[ x^{i},\dot{x}^j
\right] \left( \frac{\partial f(\mathbf{X})}{\partial x^{i}}\frac{\partial h(%
\mathbf{X})}{\partial \dot{x}^{j}}-\frac{\partial f(\mathbf{X})}{\partial 
\dot{x}^{i}}\frac{\partial h(\mathbf{X})}{\partial x^{j}}\right) .
\label{Poissonlike}
\end{equation}
We can see the relation (\ref{bracket1}) as the simple deformation of the
Poisson-like brackets introduced in (\ref{Poissonlike}). It is obvious that
the tensors $\left[ x^{i},x^{j}\right] $ and $\left[ \dot{x}^{i},\dot{x}^{j}%
\right] $ are skew sym\-me\-trics. We introduce then the following
notations 
\begin{eqnarray}
\left[ x^{i},x^{j}\right] &=&\lambda \Theta ^{ij}(\mathbf{X})\mbox{ , }%
\lambda \in \mathbb{R}  \label{xx} \\
&&  \nonumber \\
\left[ x^{i},\dot{x}^{j}\right] &=&\gamma G^{ij}(\mathbf{X})%
\mbox{
, }\gamma \in \mathbb{R}  \label{xxp} \\
&&  \nonumber \\
\left[ \dot{x}^{i},\dot{x}^{j}\right] &=&\gamma ^{\prime }\mathcal{F}{}^{ij}(%
\mathbf{X})\mbox{ , }\gamma ^{\prime }\in \mathbb{R}  \label{xpxp}
\end{eqnarray}
where $G^{ij}(\mathbf{X})$ is \textit{a priori} any $N\times N$ tensor, and
where $\Theta ^{ij}(\mathbf{X})$ and $\mathcal{F}^{ij}(\mathbf{X})$ are two $%
N\times N$ skew symmetric tensors, $\mathcal{F}^{ij}(\mathbf{X})$ being
related to the electromagnetic tensor introduced in a preceding paper
\cite{NOUS1} .

In the following we will require the property of locality 
\begin{equation}
\Theta ^{ij}(\mathbf{X})=0  \label{local}
\end{equation}
leaving the study of the $\Theta ^{ij}(\mathbf{X})\neq 0$ case for another
work \cite{NOUS5}. The property (\ref{local}) expresses the commutativity of
the internal skew-symmetric bilinear law involving positions $\mathbf{x}$ on
the manifold $\mathcal{M}$ whereas taking $\mathcal{F}^{ij}(\mathbf{X})\neq
0 $ implies noncommutativity between velocities $\dot{\mathbf{x}}$.

The property of locality (\ref{local}) and the fact that the velocity vector
has to verify the dynamical equation (\ref{dynamic}) imply the following
general expression for the Hamiltonian 
\begin{equation}
H=\frac{1}{2}mg_{ij}(\mathbf{x})\dot{x}^{i}\dot{x}^{j}+f(\mathbf{x})
\label{Hamiltonien}
\end{equation}
where we take $\lambda =1/m$ and where 
\begin{equation}
G^{ij}(\mathbf{X})=g^{ij}(\mathbf{x})
\end{equation}
is now the metric tensor def\/{}ined on the manifold $\mathcal{M}$, the
function $f(\mathbf{x})$ being any only position dependent function
belonging to $A^{0}(T(\mathcal{M}))$.

\subsection{Generalized Moyal brackets}

In this section we show that we can embedded our construction in a more
general formalism by introducing a generalization of the Moyal brackets
def\/{}ined over the tangent bundle space. Indeed we will show that the
brackets def\/{}ined by the relation (\ref{bracket1}) can be considered as
the second order expansion of the Moyal brackets where the noncommutative
parameters are played by the tensors $\Theta $, $\mathcal{F}$ and $G$
def\/{}ined in the preceding section.

Let now $f$ and $h$ be functions belonging to $A^{0}(T(\mathcal{M}))$. Then
we can def\/{}ine a Moyal star product $\star :T(\mathcal{M})\times T(%
\mathcal{M})\rightarrow T(\mathcal{M})$ such as 
\begin{equation}
f(\mathbf{X})\star h(\mathbf{X})=f(\mathbf{X})\,\,\mathrm{exp}\left( A\left( 
\mathbf{X},\mathbf{Y}\right) \right) \left. \,h(\mathbf{Y})\right| _{\mathbf{%
X}=\mathbf{Y}}
\end{equation}
where 
\begin{equation}
A\left( \mathbf{X},\mathbf{Y}\right) =\frac{1}{2}\overleftarrow{\frac{%
\partial }{\partial \mathbf{X^{\alpha }}}}a^{\alpha \beta }(\mathbf{X})%
\overrightarrow{\frac{\partial }{\partial \mathbf{Y^{\beta }}}}\quad \alpha
,\beta =1,\dots ,2N.  \label{adef}
\end{equation}
Here $\mathbf{X},\mathbf{Y}\in T(\mathcal{M})$ and the differential
operators $\overleftarrow{\frac{\partial }{\partial \mathbf{X^{\alpha }}}}$, 
$\overrightarrow{\frac{\partial }{\partial \mathbf{Y^{\beta }}}}$ are
understood to act respectively on the left and on the right side of the
expression. To this noncommutative product $\star $ we can associate a
particular commutator which is known as the Moyal brackets 
\begin{equation}
\lbrack f(\mathbf{X}),h(\mathbf{X})]_{\star }=f(\mathbf{X})\star h(\mathbf{X}%
)-h(\mathbf{X})\star f(\mathbf{X}).  \label{Moyal}
\end{equation}
Now if we give to the $2N\times 2N$ tensor $a^{\alpha \beta }$ in (\ref{adef}%
) the following antisymmetric form 
\begin{equation}
a^{\alpha \beta }(\mathbf{X})=\left( 
\begin{array}{cc}
\gamma \,\Theta (\mathbf{X}) & -\lambda G(\mathbf{X}) \\ 
\lambda G(\mathbf{X}) & \gamma ^{\prime }\mathcal{F}(\mathbf{X})
\end{array}
\right) \quad \quad \lambda ,\gamma ,\gamma ^{\prime }\in \mathbb{R}, 
\label{a}
\end{equation}
we can show that the Moyal brackets (\ref{Moyal}) are similarly to the
brackets def\/{}ined in the preceding section also a simple deformation of
the Poisson-like brackets introduced in (\ref{Poissonlike}). Indeed if we
develop the Moyal brackets up to the second order in $\lambda $, $\gamma $
and $\gamma ^{\prime }$ we obtain 
\begin{eqnarray}
\lbrack f(\mathbf{X}),h(\mathbf{X})]_{\star } &=&\{f(\mathbf{X}),h(\mathbf{X}%
)\}  \nonumber \\
&+&\gamma \,\Theta ^{ij}(\mathbf{X})\frac{\partial f(\mathbf{X})}{\partial
x^{i}}\frac{\partial h(\mathbf{X})}{\partial x^{j}}  \nonumber \\
&+&\gamma ^{\prime }\mathcal{F}^{ij}(\mathbf{X})\frac{\partial f(\mathbf{X})%
}{\partial \dot{x}^{i}}\frac{\partial h(\mathbf{X})}{\partial \dot{x}^{j}}.
\label{brackets}
\end{eqnarray}
We recall that we will limit ourselves to the case $\Theta ^{ij}(\mathbf{X})=0$
and will omit the $\star $ symbol in the following.

\section{sO(3) algebra}

\label{so3sec}

In this section we will particularly focus on the consequences of the
breaking of the sO(3) symmetry in a f\/{}lat space and in a curved space as
well. In three dimensional space the derivation of the Maxwell equations
is quite formal and we will consider only the case of magnetostatic and
electrostatic f\/ields. The time dependent f\/ields case can easily be
derived by adding a explicit time derivative to the dynamical law (\ref
{dynamic}). We do not have to add this term for the four dimensional space
case where the Maxwell equations are intrinsically Lorentz covariant,
since the f\/ields are time-dependent by construction.

\subsection{sO(3) algebra in f\/{}lat space}

In a three dimensional f\/{}lat space we have $g^{ij}(\mathbf{x})=\delta
^{ij}$ and the Hamiltonian (\ref{Hamiltonien}) of the Poisson structure
reads then 
\begin{equation}
H=\frac{1}{2}m\dot{x}^{i}\dot{x}_{i}+f(\mathbf{x}).
\end{equation}
Before considering the sO(3) algebra, let' s f\/{}irst derive the particle
equation of motion and the Maxwell f\/ield equations from our formalism.

\subsubsection{Maxwell equations}

The Jacobi identity (\ref{Jacobi}) involving position and velocity
components 
\begin{equation}
\displaystyle\frac{m}{\gamma'}
J(x^{i},\dot{x}^{j},\dot{x}^{k})=
\frac{\partial \mathcal{F}^{jk}(\mathbf{X})}{\partial \dot{x}^{i}}=0.
\end{equation}
shows that the gauge curvature is velocity independent, $\mathcal{F}^{ij}(%
\mathbf{X})\equiv F^{ij}(\mathbf{x})$. From the Jacobi identity (\ref{Jacobi}%
) involving only velocities components we derive the Bianchi equation 
\begin{equation}
\displaystyle\frac{m}{\gamma'}
J(\dot{x}^{i},\dot{x}^{j},\dot{x}^{k})=\varepsilon
^{k}{}_{ji}\displaystyle\frac{\partial F^{ij}(\mathbf{x})}{\partial x^{k}}=0
\label{Jacobi1}
\end{equation}
which, if we set $F^{ij}(\mathbf{x})=\varepsilon ^{ji}{}_{k}B^{k}(\mathbf{x}%
) $, gives the following Maxwell equation 
\begin{equation}
\mathbf{\nabla }\cdot \mathbf{B}=0.  \label{divB}
\end{equation}
Now using the dynamical equation (\ref{dynamic}) we obtain the following
equation of motion 
\begin{equation}
m\ddot{x}^{i}=m\left[ \dot{x}^{i},H\right] =qF^{ij}(\mathbf{x})%
\dot{x}_{j}+qE^{i}\left( \mathbf{x}\right)  \label{eqmotion}
\end{equation}
where 
\begin{equation}  \label{electro}
qE^{i}\left( \mathbf{x}\right) =-\frac{\partial f(\mathbf{x})}{\partial x_{i}%
}.  \label{Estatic}
\end{equation}
We have then a particle of mass $m$ and electrical charge $q$ moving in
f\/{}lat space in a presence of a magnetostatic and an electrostatic
external f\/ield. In order to get the usual form (\ref{eqmotion}) for the
equation of motion we have set $\gamma'=q/m^{2}$
in the def\/{}inition (\ref{xpxp}).

We are abble now to derive the other Maxwell equation of the f\/{}irst
group. With the dynamical equation (\ref{dynamic}) we express the time
derivative of the magnetic f\/ield 
\begin{equation}
\displaystyle\frac{dB^{i}}{dt}=\displaystyle\frac{1}{2}\varepsilon
^{i}{}_{jk}\left[ F^{jk},H\right] =\displaystyle\frac{1}{2\gamma'}
\varepsilon ^{i}{}_{jk}\left[ \left[ \dot{x}^{j},\dot{x}^{k}\right] ,H
\right]
\end{equation}
and we use the Jacobi identy (\ref{Jacobi}) to rewrite the last term of the
last equation. After some calculus we obtain 
\begin{equation}
\displaystyle\frac{dB^{i}}{dt}=-\dot{x}^{i}\mathbf{\nabla }\cdot \mathbf{B}+
\displaystyle\frac{\partial B^{i}}{\partial x_{j}}\dot{x}_{j}+\varepsilon
^{i}{}_{jk}\displaystyle\frac{\partial E^{j}}{\partial x_{k}}
\end{equation}
which using (\ref{divB}) gives the second Maxwell equation 
\begin{equation}
\displaystyle\frac{\partial \mathbf{B}}{\partial t}=-\mathbf{\nabla }\times 
\mathbf{E}=\mathbf{0}
\end{equation}
for static f\/ields and electric f\/ields deriving from any potential $f(%
\mathbf{x})$ (\ref{electro}).

As the two other Maxwell equations are not Galilean covariant they cannot
be deduced from the formalism and can be merely seen as a def\/{}inition of
the charge density and the current density. Nevertheless, as shown in the
next section the complete set of the Maxwell equations can be deduced in
the relativistic generalization.

\subsubsection{sO(3) algebra and Poincare momentum}

One of the most important symmetry in physics is naturally the spherical
symmetry corresponding to the isotropy of the physical space. This symmetry
is related to the sO(3) algebra. In the following we show that this symmetry
is broken when an electromagnetic f\/ield is applied. In order to study the
symmetry breaking of the sO(3) algebra we use the usual angular momentum $%
L^{i}=m\varepsilon ^{i}{}_{jk}x^{j}\dot{x}^{k}$ which is a constant of
motion in absence of gauge f\/ield. In fact, no electromagnetic f\/ield
implies $F^{ij}(\mathbf{x})=\left[ \dot{x}^{i},\dot{x}^{j}\right] =0$, and
the expression of the sO(3) Lie algebra with our brackets (\ref{bracket1})
gives then the standard algebra def\/{}ined in terms of the Poisson brackets
(\ref{Poissonlike}) 
\begin{equation}
\left\{ 
\begin{array}{l}
\left[ x^{i},L^{j}\right] =\left\{ x^{i},L^{j}\right\} =\varepsilon
^{ij}{}_{k}x^{k}, \\ 
\\ 
\left[ \dot{x}^{i},L^{j}\right] =\left\{ \dot{x}^{i},L^{j}\right\}
=\varepsilon ^{ij}{}_{k}\dot{x}^{k}, \\ 
\\ 
\left[ L^{i},L^{j}\right] =\left\{ L^{i},L^{j}\right\} =\varepsilon
^{ij}{}_{k}L^{k}.
\end{array}
\right.
\end{equation}
When the electromagnetic f\/ield is turned on this algebra is broken in the
following manner 
\begin{equation}
\left\{ 
\begin{array}{ccccl}
\left[ x^{i},L^{j}\right] & = & \left\{ x^{i},L^{j}\right\} & = & 
\varepsilon ^{ij}{}_{k}x^{k}, \\ 
&  &  &  &  \\ 
\left[ \dot{x}^{i},L^{j}\right] & = & \left\{ \dot{x}^{i},L^{j}\right\} & +
& \frac{q}{m}\varepsilon ^{j}{}_{kl}x^{k}{}F^{il}(\mathbf{x}) \\ 
& = & \varepsilon ^{ij}{}_{k}\dot{x}^{k} & + & \frac{q}{m}\varepsilon
^{j}{}_{kl}x^{k}{}F^{il}(\mathbf{x}), \\ 
&  &  &  &  \\ 
\left[ L^{i},L^{j}\right] & = & \left\{ L^{i},L^{j}\right\} & + & 
q\varepsilon ^{i}{}_{kl}\varepsilon ^{j}{}_{ms}x^{k}x^{m}{}F^{ls}(\mathbf{x})
\\ 
& = & \varepsilon ^{ij}{}_{k}L^{k} & + & q\varepsilon ^{i}{}_{kl}\varepsilon
^{j}{}_{ms}x^{k}x^{m}{}F^{ls}(\mathbf{x}).
\end{array}
\right.
\end{equation}
In order to restore the sO(3) algebra we introduce a new angular momentum $%
M^{i}(\mathbf{X})$ which is \textit{a priori} position and velocity
dependent. We consider then the following transformation law 
\begin{equation}
L^{i}(\mathbf{X})\rightarrow \mathcal{L}^{i}(\mathbf{X})=L^{i}(\mathbf{X}%
)+M^{i}(\mathbf{X}),
\end{equation}
and we require that this new angular momentum $\mathcal{L}^{i}$ verif\/{}ies
the usual sO(3) algebra 
\begin{equation}
\left\{ 
\begin{array}{l}
\left[ x^{i},\mathcal{L}{}^{j}\right] =\left\{ x^{i},\mathcal{L}^{j}\right\}
=\varepsilon ^{ij}{}_{k}x^{k}, \\ 
\\ 
\left[ \dot{x}^{i},{}\mathcal{L}^{j}\right] =\left\{ \dot{x}^{i},\mathcal{L}%
{}^{j}\right\} =\varepsilon ^{ij}{}_{k}\dot{x}^{k}, \\ 
\\ 
\left[ {}\mathcal{L}^{i},{}\mathcal{L}^{j}\right] =\left\{ \mathcal{L}{}^{i},%
\mathcal{L}{}^{j}\right\} =\varepsilon ^{ij}{}_{k}\mathcal{L}^{k}.
\end{array}
\right.  \label{structure}
\end{equation}
These equations (\ref{structure}) gives then three constrains on the
expression of the angular momentum $\mathcal{L}^{i}$. From the f\/{}irst
relation in (\ref{structure}) we easily deduce that $M^{i}$ is velocity
independent 
\begin{equation}
M{}^{i}(\mathbf{X})=M^{i}(\mathbf{x}),
\end{equation}
from the second relation we obtain 
\begin{equation}
\left[ \dot{x}^{i},M^{j}\right] =-\frac{1}{m}\frac{\partial M^{j}(\mathbf{x})%
}{\partial x_{i}}=-\frac{q}{m}\varepsilon ^{j}{}_{kl}x^{k}F^{il}(\mathbf{x})
\label{momentum1}
\end{equation}
and f\/{}inally the third relation gives 
\begin{equation}
M^{i}=\frac{1}{2}q\varepsilon _{jkl}x^{i}x^{k}F^{jl}(\mathbf{x})=-q\left( 
\mathbf{x}\cdot \mathbf{B}\right) x^{i}.  \label{momentum}
\end{equation}
Equations (\ref{momentum1}) and (\ref{momentum}) are compatible only if the
magnetic f\/ield $\mathbf{B}$ is the Dirac magnetic monopole f\/ield 
\begin{equation}
\mathbf{B}=\displaystyle\frac{g}{4\pi }\frac{\mathbf{x}}{\left\| \mathbf{x}%
\right\| ^{3}}.
\end{equation}
The vector $\mathbf{M}$ allowing us to restore the sO(3) symmetry (\ref
{structure}) is then the Poincar\'{e} momentum \cite{POINCARE} 
\[
\mathbf{M}=-\frac{qg}{4\pi }\frac{\mathbf{x}}{\left\| \mathbf{x}\right\| }. 
\]
already found in a preceding paper \cite{NOUS2}. The total angular momentum
is then 
\begin{equation}
\mathcal{\mathcal{L}}=\mathbf{L}-\frac{qg}{4\pi }\frac{\mathbf{x}}{\left\| 
\mathbf{x}\right\| }.
\end{equation}
This expression was initially found by Poincare in a different context
\cite{POINCARE}. Actually he was looking for an new angular momentum that would
be a constant of motion. In our framework this property is trivially
verif\/{}ied by using the dynamical relation (\ref{dynamic}).

Let us now discuss an important point. As the Dirac magnetic monopole is
located at the origin we have 
\begin{equation}
J(\dot{x}^{i},\dot{x}^{j},\dot{x}^{k})=\mathbf{%
\nabla }\cdot \mathbf{B}=g\delta ^{3}(\mathbf{x}).  \label{xpxpxpjackiw}
\end{equation}
The preservation of the sO(3) symmetry in the presence of a gauge f\/ield is
then incompatible with the requierement of the Jacobi identity at the origin
of the coordinates and we have to exclude the origin from the manifold $%
\mathcal{M}$. As the Jacobi identity is the inf\/{}initesimal statement of
associativity in the composition law of the translation group \cite{JACKIW},
the breakdown of the Jacobi identity (\ref{xpxpxpjackiw}) when $\mathbf{%
\nabla }\cdot \mathbf{B}\neq 0$ implies that f\/{}inite translations do not
associate. In usual quantum mechanics non-associativity between operators
acting on the Hilbert space can not be tolerate, one has to use the Dirac' s
quantization procedure to save associativity (\ref{xpxpxpjackiw}).


In order to consider quantum mechanics within our framework we have to
quantify as usual the totql angular momentum $\mathcal{L}$. Considering the
rest frame of the particle we have the following Dirac quantization 
\begin{equation}
\frac{gq}{4\pi }=\frac{n}{2}\hbar .
\end{equation}
Note that the Poincar\'{e} momentum is also related to the Wess-Zumino term
introduced by Witten in the case of a simple mechanical problem \cite{WITTEN}.
Indeed let' s consider a particle of mass $m$ constrained to move on a two
dimensional sphere of radius one with a spatial-temporal ref\/{}lection
symmetry. The system can be seen as a particle submitted to a strength
having the following form $qg\varepsilon _{ijk}x^{k}\dot{x}^{j}$ and which
is interpreted by Witten as a Lorentz force acting on an electric charge $q$
in interaction with a magnetic monopole of magnetic charge $g$ located at
the center of the sphere. In the quantum version of this system Witten has
recovered the Dirac quantization condition by means of topological
techniques.

\subsection{sO(3) algebra in curved space}

The Hamiltonian is now def\/{}ined in a curved space by 
\begin{equation}
H=\frac{1}{2}mg_{ij}(\mathbf{x})\dot{x}^{i}\dot{x}^{j}+f(\mathbf{x})
\end{equation}
where the metric $g_{ij}(\mathbf{x})$ is now position dependent.

\subsubsection{Maxwell equations}

The commutation relations for contravariant components are now in a curved
space 
\begin{equation}
\left\{ 
\begin{array}{lll}
\left[ x^{i},x^{j}\right] & = & 0, \\ 
&  &  \\ 
\left[ x^{i},\dot{x}^{j}\right] & = & \displaystyle\frac{1}{m}g^{ij}(\mathbf{%
x}), \\ 
&  &  \\ 
\left[ \dot{x}^{i},\dot{x}^{j}\right] & = & \displaystyle\displaystyle\frac{q%
}{m^{2}}\mathcal{F}^{ij}(\mathbf{X}).
\end{array}
\right.
\end{equation}
Then for covariant components we have 
\begin{equation}
\left\{ 
\begin{array}{lll}
\left[ x_{i},x_{j}\right] & = & 0, \\ 
&  &  \\ 
\left[ x_{i},\dot{x}_{j}\right] & = & \displaystyle\frac{1}{m}g_{ij}(\mathbf{%
x})+\displaystyle\frac{1}{m}\left( \partial _{j}g_{ik}\right) x^{k}, \\ 
&  &  \\ 
\left[ \dot{x}_{i},\dot{x}_{j}\right] & = & \displaystyle\frac{q}{m^{2}}%
\mathcal{F}_{ij}(\mathbf{X})+\displaystyle\frac{1}{m}\left( \partial
_{j}g_{ik}-\partial _{i}g_{jk}\right) \dot{x}^{k}.
\end{array}
\right.  \label{relcov}
\end{equation}
Using the Jacobi identity $J(x^{k},\dot{x}^{i},\dot{x}^{j})=0$ and using the
fact that $g^{ik}g_{kj}=\delta ^{i}{}_{j}$ we f\/{}ind the following
expression for the general gauge f\/ield, 
\begin{eqnarray}
\displaystyle\frac{q}{m^{2}}\mathcal{F}_{ij}\left( \mathbf{X}\right)
&=&g_{ik}g_{jl}\mathcal{F}^{kl}\left( \mathbf{X}\right)  \nonumber \\
&=&\displaystyle\frac{1}{m}\left( \partial _{i}g_{jk}-\partial
_{j}g_{ik}\right) \dot{x}^{k}+\displaystyle\frac{q}{m^{2}}%
F_{ij}\left( \mathbf{x}\right)
\end{eqnarray}
where $F_{ij}\left( \mathbf{x}\right) $ is velocity independent gauge
f\/ield. From the last equation in (\ref{relcov}) we easily deduce that the
commutator between covariant velocities is only position dependent 
\begin{equation}
\left[ \dot{x}_{i},\dot{x}_{j}\right] =\displaystyle\frac{q}{m^{2}}%
F_{ij}\left( \mathbf{x}\right) .
\end{equation}
Again with the dynamical equation (\ref{dynamic}) and with the above
relations of commutation we derive the equation of motion of a particle in a
curved space in the presence of a magnetostatic and electrostatic f\/ields 
\begin{equation}
m\ddot{x}^{i}=-\Gamma ^{i,jk}\dot{x}_{j}\dot{x}_{k}
+qF^{ij}(\mathbf{x})\dot{x}_{j}+qE^{i}\left( \mathbf{x}\right)
\end{equation}
where Christoffel symbols are def\/{}ined by 
\begin{equation}
\Gamma ^{i,jk}=\frac{1}{2}(-\partial ^{j}g^{ik}+\partial ^{i}g^{jk}-\partial
^{k}g^{ij}).
\end{equation}
The Jacobi identity $J\left( \dot{x}_{i},\dot{x}_{j},%
\dot{x}_{k}\right) =0$ gives directly the f{}irst Maxwell
equation of the first group 
\begin{equation}
\partial _{i}F_{jk}+\partial _{j}F_{ki}+\partial _{k}F_{ij}=0,
\end{equation}
and if we follow the same procedure as in the f{}lat space we also recover
the second Maxwell equation of the first group for the static fields 
\[
\partial _{t}B_{i}=-\varepsilon _{i}{}^{jk}\partial _{j}E_{k}=0. 
\]
As in the flat space case the second group of Maxwell equations are
considered as the def\/{}inition of the charge and the current density.

\subsubsection{sO(3) algebra and Poincare momentum}

We def\/{}ine the angular momentum in a three dimensional curved space by
the usual relations
\begin{equation}
\left\{ 
\begin{array}{l}
L_{i}=m\sqrt{g(\mathbf{x})}\varepsilon _{ijk}x^{j}\dot{x}^{k}=m
\mathrm{E}_{ijk}{}(\mathbf{x})x^{j}\dot{x}^{k} \\ 
\\ 
L^{i}=m\sqrt{g(\mathbf{x})}g^{ij}(\mathbf{x})\varepsilon _{jkl}x^{k}
\dot{x}^{l}=m\mathrm{E}^{ijk}{}(\mathbf{x})x_{j}\dot{x}_{k},
\end{array}
\right.  \label{Ldef}
\end{equation}
where $g(\mathbf{x})=\det \left( g_{ij}(\mathbf{x})\right) =\left( \det
\left( g^{ij}(\mathbf{x})\right) \right) ^{-1}$. 
Using the relation
$\partial_i g(\mathbf{x})=g(\mathbf{x})g_{jk}(\mathbf{x})
\partial_i g^{jk}(\mathbf{x})$
we easily show that the
sO(3) algebra symmetry
is broken in the following manner
\begin{equation}
\left\{ 
\begin{array}{lll}
\left[ x^{i},L^{j}\right] 
&=&
\left\{ x^{i},L^{j}\right\} =
\mathrm{E}^{ij}{}_{k}(\textbf{x})x^{k},\\ 
&&\\ 
\left[ \dot{x}^{i},L^{j}\right]
&=&
\left\{\dot{x}^{i},L^{j}\right\}
+\displaystyle\frac{q}{m}\mathrm{E}^{j}{}_{kl}
(\textbf{x})x^{k}\mathcal{F}^{il}(\textbf{X})\\ 
&=&
\mathrm{E}^{ij}{}_{k}(\textbf{x})x^{k}
-\displaystyle\frac{1}{2}\mathrm{E}^{j}{}_{kl}(\textbf{x})x^{k}\dot{x}^{l}
g_{mn}(\textbf{x})
\partial^i g^{mn}(\mathbf{x})\\
&+&\displaystyle\frac{q}{m}\mathrm{E}^{j}{}_{kl}(\textbf{x})x^{k}
\mathcal{F}^{il}(\textbf{X}), \\
&&\\ 
\left[ L^{i},L^{j}\right] 
&=&
\left\{ L^{i},L^{j}\right\}+q\mathrm{E}^{i}{}_{kl}(\textbf{x})
\mathrm{E}^{j}{}_{mn}
(\textbf{x})x^{k}x^{m}\mathcal{F}^{ln}\left(\mathbf{X}\right)\\ 
&=&
\mathrm{E}^{ij}{}_{k}(\textbf{x})L^{k}\\ 
&+&\displaystyle\frac{m}{2}
\left(g_{ab}\partial_n g^{ab}(\mathbf{x})\right)
\left(
\mathrm{E}^i{}_{kl}\mathrm{E}^j{}_m{}^n
-
\mathrm{E}^j{}_{kl}\mathrm{E}^i{}_m{}^n
\right)
x^mx^k\dot{x}^l\\
&+& q\mathrm{E}^i{}_{kl}(\textbf{x})\mathrm{E}^{j}{}_{mn}(\textbf{x})x^k
x^{m}\mathcal{F}^{ln}(\textbf{X}).
\end{array}
\right.
\end{equation}
In order to restore the sO(3) symmetry as in the f\/{}lat case we perform
the following transformation on the angular momentum 
\begin{equation}
L^{i}(\mathbf{X})\rightarrow \mathcal{{}L}^{i}(\mathbf{X})=L^{i}(\mathbf{X}%
)+M{}^{i}(\mathbf{X}),
\end{equation}
and we thus impose the constrains 
\begin{equation}\label{relwc}
\left\{ 
\begin{array}{c}
\left[ x^{i},\mathcal{L}{}^{j}\right] =\mathrm{E}{}^{ij}{}_{k}x^{k}, \\ 
\\ 
\left[ \dot{x}^{i},\mathcal{L}{}^{j}\right]
=\mathrm{E}^{ij}{}_{k}\dot{x}^{k}, \\ 
\\ 
\left[ \mathcal{L}{}^{i},\mathcal{L}{}^{j}\right]
=\mathrm{E}^{ij}{}_{k}\mathcal{L}^{k}.
\end{array}
\right.
\end{equation}
The f\/{}irst equation in (\ref{relwc}) implies that the new angular
momentum is velocity independent, $M^{i}(\mathbf{X})=M^{i}(\mathbf{x})$,
the second equation in (\ref{relwc}) gives then 
\begin{eqnarray}
\left[ \dot{x}^{i},M^{j}\right] &=&
-\frac{q}{m}\mathrm{E}^{j}{}_{lm}F^{im}x^l.
\end{eqnarray}
This last equation is similar to equation (\ref{momentum1}) found
for the flat space case, the
angular momentum $M$ is then the Poincar\'{e} momentum 
\begin{equation}
M^{i}(\mathbf{x})=\frac{1}{2}
q\mathrm{E}_{jkl}(\mathbf{x})F^{jl}(\mathbf{x)}
x^{i}x^{k}=-q\left(\mathbf{x}\cdot\mathbf{B}\right) x^{i}.
\label{MM}
\end{equation}
Still, this kind of relation (\ref{MM}) implies
a Dirac magnetic monopole field
\begin{equation}
\mathbf{B}=\frac{g}{4\pi}\frac{\mathbf{x}}{\left\|\mathbf{x}\right\|^{3}}.
\end{equation}
We have then shown that 
the sO(3) symmetry algebra in curved space is
restored by introducing the same Dirac magnetic monopole as we introduce
in the f\/{}lat
space case.

\section{Lorentz algebra in a curved space}

\label{Lorentz}

\label{lorentz}The natural extension of the previous computation is
obviously the study of the Lorentz algebra in a curved space.
In this section we consider the following Hamiltonian 
\begin{equation}
H=\frac{1}{2}mg_{\mu \upsilon }(\mathbf{x})\dot{x}^{\mu }\dot{x}^{\nu }
\end{equation}
where $g_{\mu\nu}\left(\mathbf{x}\right)$ is the metric of the Riemannian
manifold $\mathcal{M}$.

\subsection{Maxwell equations}

The simple commutation relations between positions and velocities coordinates
are
\begin{equation}
\left\{ 
\begin{array}{ccl}
\left[ x^{\mu },x^{\nu }\right]&=&0, \\ 
\\ 
\left[ x^{\mu },\dot{x}^{\nu }\right]&=&
\displaystyle\frac{1}{m}g^{\mu \nu }(\mathbf{x}), \\ 
\\ 
\left[ \dot{x}^{\mu },\dot{x}^{\nu }\right]&=&
\displaystyle\frac{q}{m^{2}}%
\mathcal{F}^{\mu \nu }(\mathbf{X}), \\ 
\\ 
\left[ \dot{x}_{\mu },\dot{x}_{\nu }\right]&=&
\displaystyle\frac{q}{m^{2}}%
F_{\mu \nu }(\mathbf{x})=\frac{q}{m^{2}}\mathcal{F}_{\mu \nu }(\mathbf{X}%
)+\left( \partial _{\nu }g_{\mu \rho }-\partial _{\mu }g_{\nu \rho }\right) 
\dot{x}^{\rho },
\end{array}
\right.
\end{equation}
from which we esaily derive the motion equation
\begin{equation}
m\ddot{x}^{\mu }=-\Gamma ^{\mu \nu \rho }\dot{x}_{\nu }
\dot{x}_{\rho }+qF^{\mu \nu }(\mathbf{x})\dot{x}_{\nu }.
\end{equation}
The Jacobi identity $J\left( \dot{x}_{\mu },\dot{x}_{\nu },%
\dot{x}_{\rho }\right) =0$ directly gives
\begin{equation}
\partial _{\mu }F_{\nu \rho }+\partial _{\nu }F_{\rho \mu }+\partial _{\rho
}F_{\mu \nu }=0
\end{equation}
which is the f\/{}irst group of the
Maxwell equations. The other Jacobi identity 
$J\left( \dot{x}_{\mu },\dot{x}^{\nu },\dot{x}_{\nu
}\right) =0$ gives us the relation
\begin{equation}
\left[ \dot{x}_{\nu },\left[ \dot{x}_{\mu },\dot{x}%
^{\nu }\right] \right] =-\left[ \dot{x}^{\nu },\left[ \dot{x}%
_{\nu },\dot{x}_{\mu }\right] \right] =\frac{q}{m^{3}}\mbox{ }
\partial^\nu F_{\nu \mu }
\end{equation}
where the quantity $\partial^\nu F_{\nu \mu }$
is {\it a priori} different from zero. We def\/{}ine then
the second group of the generalized Maxwell equations by
\begin{equation}
\left\{ 
\begin{array}{c}
\partial ^{\nu }F_{\nu \mu }=0\mbox{ , for the vacuum} \\ 
\\ 
\partial ^{\nu }F_{\nu \mu }=j_{\mu }
\mbox{ , for a medium with a current
density }j^{\alpha }
\end{array}
\right\}.
\end{equation}

\subsection{Lorentz algebra and Poincare momentum}

It is convenient to def\/{}ine the angular quadrimomentum as
\begin{equation}
L_{\mu \nu }=m\left( x_{\mu }\dot{x}_{\nu}-x_{\nu}\dot{x}_{\mu }\right) ,
\end{equation}
which gives a deformed Lorentz algebra with the following structure
\begin{equation}
\left\{ 
\begin{array}{l}
\left[ x_{\mu },L_{\rho \sigma }\right] =\left\{ x_{\mu },L_{\rho \sigma
}\right\} \\ 
=g_{\mu \sigma }(\mathbf{x})x_{\rho }-g_{\mu \rho }(\mathbf{x})x_{\sigma
}+x_{\rho }x^{\lambda }\partial_\sigma g_{\mu \lambda }(\mathbf{x})
-x_{\sigma }x^{\lambda }\partial_\rho g_{\mu \lambda }(\mathbf{x}), \\ 
\\
\left[ \dot{x}_{\mu },L_{\rho \sigma }\right] =\left\{ \dot{x_{\mu }}%
,L_{\rho \sigma }\right\}
+\displaystyle\frac{q}{m}(F_{\mu \sigma }(\mathbf{x})\dot{%
x_{\rho }}-F_{\mu \rho }(\mathbf{x})\dot{x_{\sigma }}) \\ 
=g_{\mu \sigma }(\mathbf{x})\dot{x_{\rho }}-g_{\mu \rho }(\mathbf{x})
\dot{x_{\sigma }}+\dot{x}_{\rho }x^{\lambda }
\partial_\sigma g_{\mu \lambda }(\mathbf{x})
-\dot{x}_{\sigma }x^{\lambda }
\partial_\rho g_{\mu \lambda }(\mathbf{x})\\ 
+\displaystyle\frac{q}{m}(F_{\mu \sigma }(\mathbf{x})\dot{x_{\rho }}
-F_{\mu \rho }(\mathbf{x})\dot{x_{\sigma }}), \\ 
\\
\left[ L_{\mu \nu },L_{\rho \sigma }\right] =\left\{ L_{\mu \nu },L_{\rho
\sigma }\right\} \\ 
+q(x_{\mu }x_{\rho }F_{\nu \sigma }(\mathbf{x})-x_{\nu }x_{\rho }
F_{\mu\sigma }(\mathbf{x})
+x_{\mu }x_{\sigma }F_{\rho \nu }(\mathbf{x})-x_{\nu
}x_{\sigma }F_{\rho \mu }(\mathbf{x})) \\ 
=g_{\mu \rho }(\mathbf{x})L_{\nu \sigma }-g_{\nu \rho }(\mathbf{x})L_{\mu
\sigma }+g_{\mu \sigma }(\mathbf{x})L_{\rho \nu }-g_{\nu \sigma }
(\mathbf{x})L_{\rho \mu } \\ 
+m\left( x_{\rho }\dot{x}_{\nu }x^{\lambda }
\partial_\mu g_{\lambda \sigma}(\mathbf{x})
-x_{\nu }\dot{x}_{\rho }x^{\lambda }
\partial_\sigma g_{\lambda \mu }(\mathbf{x})
+x_{\mu }\dot{x}_{\rho }x^{\lambda }
\partial_\sigma g_{\lambda \mu }(\mathbf{x})\right. \\ 
\left.-x_{\rho }\dot{x}_{\mu }x^{\lambda }
\partial_\mu g_{\lambda \sigma}(\mathbf{x})
+x_{\nu }\dot{x}_{\sigma }x^{\lambda }
\partial_\mu g_{\lambda \rho }(\mathbf{x})
-x_{\sigma }\dot{x}_{\nu }x^{\lambda }
\partial_\rho g_{\lambda \mu }(\mathbf{x})\right. \\ 
\left. +x_{\sigma }\dot{x}_{\mu }x^{\lambda }
\partial_\rho g_{\lambda \nu }(\mathbf{x})
-x_{\mu }\dot{x}_{\sigma }x^{\lambda }
\partial_\nu g_{\lambda \rho }(\mathbf{x})\right) \\ 
+q(x_{\mu }x_{\rho }F_{\nu \sigma }(\mathbf{x})-x_{\nu }x_{\rho }F_{\mu
\sigma }(\mathbf{x})+x_{\mu }x_{\sigma }F_{\rho \nu }(\mathbf{x})-x_{\nu
}x_{\sigma }F_{\rho \mu }(\mathbf{x})).
\end{array}
\right.
\end{equation}

We apply here the same scheme used for the sO(3) algebra, we restore
the Lorentz symmetry by using the following
angular quadrimomentum transformation law 
\begin{equation}
L_{\mu \nu }(\mathbf{X})\rightarrow\mathcal{L}_{\mu \nu }(\mathbf{X})
=L_{\mu \nu }(\mathbf{X})
+M_{\mu\nu }(\mathbf{X}),
\end{equation}
and by requiring the usual structure
\begin{equation}
\left\{ 
\begin{array}{lllll}
\left[ x_{\mu },\mathcal{L}_{\rho \sigma }\right] & = & \left\{ x_{\mu },
\mathcal{L}_{\rho \sigma }\right\}&=&g_{\mu \sigma }x_{\rho }-g_{\mu \rho
}x_{\sigma },\label{lie1} \\ 
&  &  &&\\ 
\left[ \dot{x}_{\mu },\mathcal{L}_{\rho \sigma }\right] & = & \left\{ \dot{x}%
_{\mu },\mathcal{L}_{\rho \sigma }\right\}&=&g_{\mu \sigma }\dot{x}_{\rho
}-g_{\mu \rho }\dot{x}_{\sigma },\label{lie3} \\ 
&  &  &&\\ 
\left[ \mathcal{L}_{\mu \nu },\mathcal{L}_{\rho \sigma }\right] & = & 
\left\{\mathcal{L}_{\mu \nu },\mathcal{L}_{\rho \sigma }\right\}&=&
g_{\mu \rho}
\mathcal{L}_{\upsilon \sigma }
-g_{\nu \rho }
\mathcal{L}_{\mu \sigma }
+g_{\mu \sigma }
\mathcal{L}_{\rho\nu }
-g_{\nu \sigma }
\mathcal{L}_{\rho \mu }.
\end{array}
\right.  \label{lie5}
\end{equation}
From (\ref{lie1}) we easily deduce that the quadrimomentum
$M_{\mu\nu}$ is only position dependent,
$M_{\mu \nu }(\mathbf{X})=M_{\mu \nu }(\mathbf{x})$.
Then the equation (\ref{lie3}) also gives 
\begin{equation}
\left[ \dot{x}_{\mu },M_{\rho \sigma }\right] =\frac{q}{m}(F_{\mu \sigma
}x_{\rho }-F_{\mu \rho }x_{\sigma }).  \label{neufter}
\end{equation}
This result (\ref{neufter}) with the third relation given in 
(\ref{lie5}) give us the following relation 
\begin{equation}
g_{\mu \rho }M_{\nu \sigma }-g_{\nu \rho }M_{\mu \sigma }+g_{\mu \sigma
}M_{\rho \nu }-g_{\nu \sigma }M_{\rho \mu }=q(F_{\nu \sigma }x_{\mu }x_{\rho
}-F_{\mu \sigma }x_{\nu }x_{\rho }+F_{\rho \nu }x_{\mu }x_{\sigma }-F_{\rho
\mu }x_{\nu }x_{\sigma }).  \label{neuf}
\end{equation}
which will define the quadrimomentum $M_{\mu\nu}$.

First, let us consider
the case $\nu =\sigma =i$, where $i=1,2,3$, and with a
sum over $i$. Equation (\ref{neuf}) becomes  then
\begin{equation}
-g^{i}{}_{\rho }M_{\mu i}(\mathbf{x})+g_{\mu }{}^{i}M_{\rho i}(\mathbf{x)}
-3M_{\rho \mu }(\mathbf{x})=q(-F_{\mu i}(\mathbf{x})x^{i}x_{\rho }+F_{\rho
i}(\mathbf{x})x_{\mu }x^{i}-F_{\rho \mu }(\mathbf{x})\mathbf{x}^{2}), 
\label{neufbis}
\end{equation}
Now setting $\rho =j$ and $\mu =k$, we obtain 
\begin{equation}
M_{ij}=q(F_{ij}x^{k}x_{k}-F{}_{jk}x^{k}x_{i}-F_{ki}{}x^{k}x_{j})
\label{Equation Moment}
\end{equation}
which is nothing more than the generalization of the
previously found equations (\ref{momentum}) and (\ref{MM}).
$M_{ij}$ is then the quadrimomentum related to the previously
found Poincar\'e momentum $M_i$.
Indeed using the def\/{}inition of the quadrimomentum
$M_{i}=\varepsilon {}_{i}{}^{jk}M_{jk}$, we retrieve 
for the spatial degrees of freedom ($i=1,2,3$) the Poincar\'e momentum
\begin{equation}
\mathbf{M}=-q(\mathbf{x}\cdot\mathbf{B})
\mathbf{x}.  \label{ChampMagnetique}
\end{equation}
Using now the equations (\ref{ChampMagnetique}) and (\ref{neufter}) we
obtain the set of equations 
\begin{equation}
\left\{ 
\begin{array}{l}
x_{i}B_{j}+x_{j}B_{i}=-x_{j}x^{k}\partial_i B_{k}, \\ 
\\ 
F_{0j}x_{i}-F_{0i}x_{j}=
\left( \mathbf{x}\times\mathbf{E}\right) _{k\neq i,j}=0,
\end{array}
\right.  \label{relation}
\end{equation}
whose solutions are radial vector f\/ields centered at the origin 
\begin{equation}
\left\{ 
\begin{array}{l}
\mathbf{B}=\displaystyle\frac{g}{4\pi }
\displaystyle\frac{\mathbf{x}}{\left\|\mathbf{x}\right\|^{3}},\\ 
\\ 
\mathbf{E}=q'
f(\mathbf{x})\mathbf{x}.
\end{array}
\right.  \label{EB}
\end{equation}
It is straightforward
to see that our results are still valid for a f\/{}lat 
quadridimensional space.
We have then shown that the Lorentz symmetry in a curved and in a
f\/{}lat space is restored
if the magnetic f\/ield is the Dirac monopole magnetic field
and if the electric f\/ield is radial.

Consider now the ''boost'' part of (\ref{neuf}). For $\rho =0$, and $\mu =j$%
, equation (\ref{neufbis}) corresponds to the temporal components of the
Poincar\'{e} tensor 
\begin{equation}
M_{0j}=q(-F_{ji}x^{i}x_{0}-F_{0i}x_{j}x^{i}+F_{0j}\mathbf{x}^{2}).
\end{equation}
This relation can also be written
\begin{equation}
M_{0j}=q\left[ -\left( \mathbf{x}\times \mathbf{B}\right)_{j}x_{0}
-\left( \mathbf{x}\cdot\mathbf{E}\right)
x_{j}+\mathbf{x}^{2}E_{j}\right]
\end{equation}
which for the solution (\ref{EB}) gives the result 
\[
M_{0j}=0.
\]
The temporal component of the generalized angular momentum $M_{\mu\nu}$
which restore the
Lorentz symmetry mixes the electric and the magnetic f\/ields in such way that
it is equal to zero, whereas the spatial components of $M_{\mu\nu}$
are only magnetic
f\/ield dependant and correspond to the usual Poincare momentum components.
It is
important to precise that the equations (\ref{EB}) implies
that the source of the
electromagnetic f\/ield is created by a Schwinger dyon of magnetic charge $g$
and electric charge $q'$.

In the next section we extend our model by adding to the electromagnetic
$F^{\mu\nu}$ tensor its dual $^* F^{\mu\nu}$
in a curved space and we deduce in the frame of the
gravitoelectromagnetism a quantization of the dyon's mass.

\subsection{Gravitoelectromagnetism}

\label{gravito}

In order to interpret the experimental tests of gravitation theories, the
Pa\-ra\-me\-tri\-zed
Post-Newtonian formalism (PPN) is often used \cite{WHEELER},
where the limit of low velocities and small stresses is taken. In this
formalism, gravity is described by a general type metric containing
dimensionless constants call PPN-parameters, which are powerful tools in
theoretical astrophysics. This formalism was applied by Braginski \textit{et
al} \cite{BRAGINSKI} to propose laboratory experiments to test relativistic
gravity and in particular to study gravitoelectromagnetism. They analyzed
magnetic and electric type gravity using a truncated and rewritten version
of the PPN formalism by deleting certain parameters not present in general
relativity and all gravitational non linearities. In a theoretical paper
\cite{MASHHOON}
Mashhoon has considered several important quantities relative to
this theory, like f\/ield equations, gravitational Larmor precession or
stress-energy tensor. He introduced gravitoelectromagnetism which is based
upon the formal analogy between gravitational Newton potential and electric
Coulomb potential. A long time ago, Holzmuller \cite{HOLZMULLER} and
Tisserand \cite{TISSERAND} have already postulated gravitational
electromagnetic components for the gravitational inf\/{}luence of the sun on
the motion of planets. More recently, Mashhoon \cite{MASHHOON} has considered a
particle of inertial mass $m$ which has also a gravitoelectric charge $%
q_{E}=-m$ and gravitomagnetic ch$\arg $e $q_{M}=-2m$, the numerical factor
2 coming from the spin character of the gravitational f\/ield. In the
f\/{}inal part of this work we apply our formalism to this last idea.

Suppose that gravitation creates a gravitoelectromagnetic f\/ield 
characterized by $F_{\mu\nu }(\mathbf{x})$ and
$^{\ast }\!F_{\mu \nu }(\mathbf{x})$ where the symbol $^*$ stands here for
the Hodge duality. We then have
\begin{equation}
\left[ \dot{x}_{\mu },\dot{x}_{\nu }\right] =-\frac{1}{m^{2}}(qF_{\mu \nu }
(\mathbf{x})+g^{\ast }\!F_{\mu \nu }(\mathbf{x}))
\end{equation}
where $q$ and $g$ are respectively the gravitoelectric and the gravitomagnetic
charge of a Schwinger dyon moving in this gravitoelectromagnetic f\/ield.
At the end of
this section we will choose like Mashoon an explicit relation between
the charges $q$ and $g$ and the inertial mass $m$. By a direct application of
our formalism developped in section \ref{Lorentz}
the equation of motion of our dyon particle
is obtained
\begin{equation}
m\ddot{x}^{\mu }=-\Gamma ^{\mu \nu \rho }\dot{x}_{\nu }%
\dot{x}_{\rho }+\left( qF^{\mu \nu }(\mathbf{x})+g^{*}\!F_{\mu \nu }(%
\mathbf{x})\right) \dot{x}_{\nu }.
\end{equation}

In order to restore the Lorentz symmetry in a curved space we introduce the
generalized angular momentum now expressed in terms of the electomagnetic
f\/ield and its dual. For the spatial components we have the following
equation
\begin{eqnarray}
M{}_{ij} &=&q(F_{ij}x^{k}x_{k}-F_{jk}x^{k}x_{i}-F_{ki}x^{k}x_{j})  \nonumber
\\
&&+g(^{\ast }\!F_{ij}x^{k}x_{k}-^{\ast }\!F_{jk}x^{k}x_{i}-^{\ast
}\!F_{ki}x^{k}x_{j}).
\end{eqnarray}
which shows that the new angular momentum is the sum of two contributions, a
gravitomagnetic one and a gravitoelectric one 
\begin{equation}
\mathbf{M}=\mathbf{M_{m}}+\mathbf{M_{e}},
\label{vingt deux}
\end{equation}
where 
\begin{equation}
\left\{ 
\begin{array}{ccr}
\mathbf{M_{m}}&=&-q(\mathbf{x}\cdot\mathbf{B})\mathbf{x}, \\ 
&&\\ 
\mathbf{M_{e}}&=&g(\mathbf{x}\cdot\mathbf{E})\mathbf{x}
\end{array}
\right.
\end{equation}
are respectively 
the gravitomagnetic and gravitoelectric angular momenta.
Introducing the notation
$\mathbf{P}=
q'\mathbf{B}-g'\mathbf{E}$
we can show that the Lorentz symmetry is restored by the
Poincar\'e-like angular momentum
\begin{equation}
\mathbf{M}=-(\mathbf{x}\cdot\mathbf{P})\mathbf{x}
\label{M}
\end{equation}
where $\mathbf{P}$ has the Dirac-like form
\begin{equation}
\mathbf{P}\sim \frac{\mathbf{x}}{4\pi\left\|\mathbf{x}\right\|^{3}}.
\end{equation}
A possible choice for the electromagnetic f\/ield is then to choose a dyon
source responsible for the Dirac and the Coulomb monopole fields, 
\begin{equation}
\left\{ 
\begin{array}{ccr}
\mathbf{B}&=&
\displaystyle\frac{g'}{4\pi }\frac{\mathbf{x}}{\left\|\mathbf{x}\right\|} \\ 
&&\\ 
\mathbf{E}&=&
-\displaystyle\frac{q'}{4\pi }\frac{\mathbf{x}}{\left\|\mathbf{x}\right\|}
\end{array}
\right.
\end{equation}
so that 
\begin{equation}
\mathbf{P}=\left( q'g+g'q\right)
\displaystyle\frac{\mathbf{x}}{4\pi \left\|\mathbf{x}\right\|^{3}}.  \label{P}
\end{equation}
Consequently we have in presence a gravitoelectromagnetic dyon
(characterized by its mass $m$ and its charges  $q$ and $g$)
moving in a gravitoelectromagnetic
monopole field creates by a dyon (characterized by 
its mass $m'$ and its charges $q'$ and $g'$).

From the quantization of (\ref{M}) with $P$ given by (\ref{P})
we deduce the following relation
\begin{equation}
\frac{q'g+g'q}{4\pi }=\frac{n\hbar }{2}.
\end{equation}
We postulate as in Ref. \cite{MASHHOON} the following relations between the
gravitoelectromagnetic charges and the inertial masses 
\begin{equation}
\left\{ 
\begin{array}{ccc}
q & = & a\sqrt{G}m, \\ 
&  &  \\ 
g & = & b\sqrt{G}m, \\ 
&  &  \\ 
q' & = & a\sqrt{G}m', \\ 
&  &  \\ 
g' & = & b\sqrt{G}m',
\end{array}
\right.
\end{equation}
where $a$ and $b$ are two constants and $G$ is the gravitational constant.
We have also $\displaystyle\frac{q_{M}}{q_{E}}=\frac{q_{M}'}{q_{E}'}
=\frac{b}{a}=s$ which is the Mashhoon's relation between electric and magnetic
charges and the spin of the gauge boson interaction. In the
gravitoelectromagnetic theory we naturally choose $s=2$, we then deduce
the following mass condition for the dyons
\begin{equation}
m\,m'=nA\frac{hc}{G}=nAM_{P}^{2}
\end{equation}
where $A$ is a dimensionless constant and $n$ is an integer number in the
Schwinger formalism (bosonic spectrum) and half integer number in the Dirac
formalism (fermionic and bosonic spectrum), $M_{P}$ being the Planck mass.

\section{Conclusion}

\label{conclusion}
In this paper we have introduced a Poisson structure with
a dynamic def\/{}ined through a covariant Hamiltonian. Our formalism could
also be expressed in terms of a generalization of the Moyal brackets
def\/{}ined on the tangent bundle space. In noncommutative theories the
parameter $\Theta ^{ij}$\ expresses the noncommutativity of positions,
whereas in our construction it is the electromagnetic f\/ield $F^{ij}\left(
x\right) $ which induces the noncomutativity of the velocities. Our aim was
to f\/{}ind the generalized angular momentum which enable us to restore the
Lie algebra symmetry ($sO(3)$ and Lorentz) of the angular momentum which is
broken by the electromagnetic f\/ield, i.e. by the noncommutativity of the
velocities. The solution is the Poincar\'{e} angular momentum in the
f\/{}lat space case as well as in the curved space case. The formalism was
applied in the framework of the gravitoelectromagnetism where it was shown
that the Dirac and the Coulomb monopoles allow to build both a magnetic and
a electric Poincare like angular momentum which restore the Lorentz algebra
symmetry in a curved space. The quantization of the total angular momentum
and the Mashoon' s relation between the gravitoelectromagnetic charges and
the inertial masses leads to a qualitative condition on the mass spectrum.

It would be interesting to extend our approach in the context of the
noncommutativity theory where $\Theta ^{ij}\neq 0$, work in that direction
is in progress \cite{NOUS5}.

\textbf{Acknowledgment:} A.B. would like to thank Patrice P\'{e}rez for
helpful discussions.

\end{document}